\documentclass[]{spie}  

\usepackage[]{graphicx}

\title{A Bayesian approach to filter design: detection of compact sources} 

\author{Marcos L\'opez-Caniego\supit{a,b}, Diego Herranz\supit{c},
  Rita Bel\'en Barreiro\supit{a} and Jos\'e Luis Sanz\supit{a}
\skiplinehalf
\supit{a} Instituto de F\'\i sica de Cantabria, Avda. de los Castros
  s/n, 39005 Santander, Spain; \\
\supit{b} Dpto. F\'\i sica Moderna, Avda. de los Castros
  s/n, 39005 Santander, Spain; \\
\supit{c}ISTI-CNR, via G. Moruzzi 1, 56124 Pisa, Italy}

\authorinfo{Further author information: (Send correspondence to MLC)
\\ MLC: E-mail: caniego@ifca.unican.es}

\begin{document} 

\maketitle 

\begin{abstract}
We consider filters for the detection and extraction of compact
sources on a background. We make a one-dimensional treatment (though a
generalization to two or more dimensions is possible) assuming that
the sources have a Gaussian profile whereas the background is modeled by
an homogeneous and isotropic Gaussian random field, characterized by a
scale-free power spectrum. Local peak detection is used after
filtering. Then, a Bayesian Generalized Neyman-Pearson test is
used to define the region of acceptance that includes not only the
amplification but also the curvature of the sources and the a priori
probability distribution function of  the sources. We search for an 
optimal filter between a family of Matched-type filters (MTF) modifying the 
filtering scale such  that it gives the maximum number of real 
detections once fixed the number density of spurious sources. 
We have performed numerical simulations to test theoretical  ideas.
\end{abstract}



\section{INTRODUCTION}
\label{sect:intro}  

The detection of compact signals (sources) embedded in a background is 
a recurrent problem in many fields of science. Some common 
examples in Astronomy are the separation of  
individual stars in a crowded optical image, the identification of
local features (lines) in noisy one-dimensional spectra
or the detection of faint extragalactic objects in microwave frequencies. 

Regarding the detection of point sources on maps of the cosmic
microwave background radiation (CMB), 
several techniques based on different linear filters have been proposed 
in the literature: the Mexican Hat Wavelet (MHW Cay\'on et al.\cite{cayon00}, 
Vielva et al.\cite{vielva01a,vielva01b}),
the classic \emph{matched} filter (MF, Tegmark and
de Oliveira-Costa\cite{tegmark98}), the Adaptive Top Hat
Filter (ATHF, Chiang et al.\cite{chiang02}) and the scale-adaptive filter (SAF,
Sanz et al.\cite{sanz01}, Herranz et
al.\cite{herranz02a,herranz02b,herranz02c}). 
A certain deal of controversy has appeared about which one, if any, of the
previous
filters is \emph{optimal} for the detection of point sources in CMB data.

In order to answer that question it is necessary to consider first a more
fundamental
issue, the concept of \emph{detection} itself. 
The detection process can be posed as follows: given an observation, the
problem is to \emph{decide} whether or not a certain signal was present at the
input of the
receiver. The decision is not obvious since the observation is corrupted by 
a random process that we call `noise' or `background'.

 Formally, the \emph{decision}
is performed by choosing between two complementary hypotheses: that the 
observed data is originated by the background alone 
(\emph{null hypothesis}), and the hypothesis that
the observation corresponds to a combination of the background and the signal.
To decide, the detector should use the totality of the available information in
terms of
the probabilities of both hypotheses given the data. The 
\emph{decision device} separates the space $\mathcal{R}$ of all possible 
observations in two disjoint subspaces, $\mathcal{R}_*$ and $\mathcal{R}_-$, 
so that if an observation 
$y \in \mathcal{R}_-$ the null hypothesis is accepted, and if $y \in
\mathcal{R}_*$ 
the null hypothesis is rejected, that is, a source is `detected'.
Hence, we will call any generic decision device 
of this type a \emph{detector}.

Any detector can produce two kinds of errors: on the one hand, 
it can produce a \emph{false alarm} or \emph{spurious detection} when 
an observation in which no source was present is assigned to the subspace
$\mathcal{R}_*$.
The probability of this kind of error depends on the statistical properties of
the background
and the choice of the detector.
On the other hand, a signal that is present in the observation can be missed by
the
detector (i. e. the observation is wrongly assigned to the
subspace $\mathcal{R}_-$). This error is often referred as \emph{false
dismissal}.
The probability of false dismissal depends on the statistical properties
of the background, the choice of the detector and the 
properties of the signal (for example, its intensity).
In general, it is not possible to decrease the incidence of both types of error
at the same time: one of them can be reduced at expense of increasing the other.
The goodness of a given detector must be established
by taking into account the balance between these two types of error.

The most simple example of detector, and one that has been
exhaustively used in Astronomy, is \emph{thresholding}.
Thresholding considers that the space $\mathcal{R}$ of observations
consists of all the possible values of the measured intensity $\xi$ (in the case
of an astronomical image) and subdivides this space into two simple regions
$\mathcal{R_-} \equiv \{ \xi \in \mathcal{R} : \xi < \xi_* \}$ and
$\mathcal{R_*} \equiv \{ \xi \in \mathcal{R} : \xi \geq \xi_* \}$.
The value $\xi_*$ is an arbitrarily chosen \emph{threshold}
that is often expressed as a number of times the standard deviation of the
background, $\xi_* = \nu_* \sigma_0$.
Thresholding works on the assumption that the probability of 
finding a value of $\xi$ due to the background decreases as the 
value of $\xi$ increases. 
In the case of a Gaussian background, this assumption has a very
precise meaning and it allows us to straightforwardly control the 
probability of occurrence of spurious detections simply by 
setting a large enough threshold.
However, this may lead to a very high probability of
false dismissals.

Unfortunately, in many cases the sources are very faint and this makes very
difficult to
detect them: a high threshold means that the number of detections will be
very small. Here is where \emph{filtering} enters in scene. 
The role of filtering is to transform the data in such a way that
a detector can perform better than before filtering. For example, a filter 
can be designed in order
to reduce the fluctuations of the background so that we can safely use lower
detection thresholds and, hopefully, increase the number of detections without
increasing the number of spurious detections. We remark that \emph{a filter is
not a 
detector}: the decision device we call `detector' can be applied 
after the application of any imaginable filter, or even no filter at all, while
the use of any filter without a posterior detection criterion means nothing.
However, the two different steps in the process (filtering and detection) are
not independent. 
In the thresholding example, the use of a filter that 
cancels most of the fluctuations in the background allows us to 
change the detection threshold from its original high value to a lower one.
Given an adopted detector and a background, it is licit to ask which is the
filter that creates the most favorables conditions in the filtered background
for the
detector to perform. In other words, the `optimality' of a filter for detection
depends
on the type of detector chosen which, in turn, depends on the specific
\emph{goal}
the observer has in mind: in certain cases the observer will accept a relatively
large 
number of spurious detections in order to have a large number of true targets,
whereas
in other cases it could be more important to be certain that the detections are
all
of them reliable, and so on.

For example, let us consider that we have chosen thresholding 
as our detection device. In that case, it has been shown that the optimal linear filter
is the matched filter. It produces the maximum amplification of the signal
with respect to the background fluctuations, so that the threshold 
for a given probability of spurious detections is minimum, allowing the 
thresholding detector to find more sources than would be detected if we
filtered with any other filter.  
A sub-optimal approach, using only thresholding, is to select a priori a filter and adapt its scale
in order to produce a maximum amplification in a given background.
Such is the case of the Mexican Hat Wavelet at the optimum scale (MHO,
Vielva et al.\cite{vielva01a,vielva01b})
and the Adaptive Top Hat Filter (ATHF, Chiang et al.\cite{chiang02}).

Thresholding has a number of advantages, among them the facts that it is 
straightforward, 
it has an obvious meaning in the case of Gaussian backgrounds,
and it has been successfully used for many years 
in many fields of science. It, however, does not use all the available
information contained in the data to perform decisions. The question is then:
is it possible to devise a detector that uses additional information 
apart from mere intensities and that produces better results than thresholding?
And, if so, which is the filter that optimizes the performance of such a
detector?

Let us focus on the case of one-dimensional data (such as stellar spectra, or
time-ordered
scannings of the sky in CMB experiments) and linear filters. 
Data in a one-dimensional array is entirely described by two quantities,
namely the position in the array (corresponding to the spatial or temporal
coordinate,
for simplicity we will refer it as \emph{spatial} information)
and the value (intensity) at each position. Thresholding uses only the
intensity 
distribution
to make the decision.
Clearly, the inclusion of  spatial information in a detector
should be useful. For example, it could help to distinguish 
the sources from fluctuations in the background with similar scale but a
different
shape. 
A full description of this `spatial information' should include the probability
distribution of
events (both due to background and sources) in space, with all its infinite
moments. We will
somewhat relax this demand of information assuming that the background is
homogeneous
and isotropic, 
and asking at each point for some information about the shape of the
sources and the autocorrelation of the background (for example, the
curvature of the peaks).

In fact, even a simple filtering-and-thresholding scheme uses implicitly some
degree of 
spatial information. Both the MHO and the ATHF adapt to the scale at which the 
contrast between sources and background produces the maximum amplification.
The MF includes as well the information on the profile of the sources in order
to
amplify the structures whose shape correlates with the shape of the sources. SAF
goes a
step further in constraining additionally the scale of the filter. Moreover, 
in most cases the detection is performed not in all the points of the data but
only
in the peaks, that is, \emph{in those points where the curvature is positive}. 

In a recent work, Barreiro et al.\cite{barreiro03} propose a detection
criterion based on the Neyman-Pearson decision rule 
that uses the information of both the intensities $\xi$ and the
curvatures $\kappa$ of the peaks in a data set. In that work
the performances of several filters (SAF, MF and MHW) 
is compared in terms of their \emph{reliability},
defined as the ratio between the probability density of true detections
over the probability density of spurious detections. They find that,
on the basis of this quantity, the choice of the optimal filter
depends on the statistical properties of the background.
For the case of backgrounds that can be described with a power spectrum of the
form
$P(q) \propto q^{-\gamma}$, the SAF outperforms the other two filters for the
case $1 < \gamma \leq 1.6$, whereas in the range $0 \leq \gamma \leq 1$ the
MF
is the most reliable. The MHW is the most reliable filter in this sense
when $\gamma > 1.6$.

The reliability, defined in the previous sense, could not be 
a valid measure of performance of the
filters, since it favors a situation in which the number of the detections is 
very low in order to keep a `safe' number of spurious detections.
A different approach can be used in which the number density of spurious
detections (or, alternatively, of true detections)
is fixed for all the filters, and then the number density of true detections
(or spurious detections) is compared for all the filters. 
 In this work, we first clearly define the \emph{goal} of our experiment: fixed
a certain number density of spurious detections, to obtain the maximum 
possible true detections from the data. Note that this goal 
is not universal: in other applications, an observer may desire to
work on the basis of the reliability described in Barreiro et
al.\cite{barreiro03}, 
or define its own requirements. 
Once fixed the goal, we will develop a detection criterion that 
corresponds to a Bayesian Generalized Neyman-Pearson test. 
Then, a particular filter from the family of matched-type filters (MTF) 
that optimizes the performance of the 
detector under the optimality conditions set by our goal will be obtained.
Finally, the performance of the optimal MTF will be compared with the standard MF using simulations.

The overview of this paper is as follows:
In section 2, we introduce two useful quantities: number of maxima in a 
Gaussian background in the absence and presence of a local source. 
In section 3, we introduce the detection problem and define the region of
acceptance. In section 4, we obtain different analytical
and numerical results regarding point sources and scale-free
background spectra  
and compare the performance of a new family of matched-type filters. 
In section 5, we describe the numerical simulations performed to test some 
theoretical aspects and give the main results. Finally, in section 6, 
we summarize the main results and applications of this paper.


\section{BACKGROUND PEAKS AND COMPACT SOURCES}

\subsection{The background}

Let us assume a 1D background (e. g. one-dimensional scan on the
celestial sphere or time 
ordered data set) represented by a Gaussian random field $\xi (x)$
with average value  
$\langle \xi (x)\rangle = 0$ and power spectrum $P(q), \ q\equiv |Q|$: 
$\langle \xi (Q)\xi^* (Q')\rangle = P(q)\delta_D (q - q')$, where
$\xi (Q)$ is the Fourier transform of $\xi (x)$ and $\delta_D$ is the 1D Dirac 
distribution. 
The distribution of
maxima was studied by Rice\cite{rice54} in a pioneer article. The expected
number density of 
maxima per intervals $(x, x + dx)$, $(\nu ,\nu + d\nu )$ and $(\kappa
,\kappa + d\kappa )$ is given by
\begin{equation} 
n_b(\nu ,\kappa )  = \frac{n_b\,\kappa}{\sqrt{2\pi (1-\rho^2)}} 
e^{- \frac{\nu^2 + \kappa^2 - 2\rho \nu \kappa}{2(1 - \rho^2)}}, 
\label{nbackground}
\end{equation}
being $n_b$ the expected total number density of maxima (i. e. number of 
maxima per unit interval $dx$)
\begin{equation}  
n_b \equiv \frac{1}{2\pi \theta_m},\ \ \ 
\nu \equiv \frac{\xi}{\sigma_0},\ \ \ 
\kappa \equiv \frac{-\xi^{\prime \prime}}{\sigma_2}, \\ 
\end{equation}

\begin{equation}
\theta_m \equiv \frac{\sigma_1}{\sigma_2},\ \ \ 
\rho \equiv \frac {\sigma_1^2}{\sigma_0 \sigma_2} = \frac{\theta_m}{\theta_c},\
\ \ 
\theta_c \equiv \frac{\sigma_0}{\sigma_1}, \nonumber 
\end{equation}
where $\nu \in (-\infty ,\infty )$ and $\kappa \in (0,\infty )$ represent the 
normalized field and curvature, respectively. $\sigma_n^2$ is the
moment of order $2n$ associated to the field. $\theta_c ,\theta_m $
are the coherence scale of the field and maxima, respectively. 

If the original field is linear-filtered with a circularly-symmetric 
filter $\Psi (x; R, b)$, dependent on
$2$ parameters ($R$ defines a scaling whereas $b$ defines a translation)
\begin{equation}
\Psi (x; R, b) = \frac{1}{R}\psi \left(\frac{|x - b|}{R}\right),
\end{equation}
we define the filtered field as
\begin{equation}
w(R, b) = \int dx\,\xi (x)\Psi (x; R, b).
\end{equation}
Then, the moment of order $n$ of the linearly-filtered field is
\begin{equation}
\sigma_n^2 \equiv 2\int_0^{\infty} dq\,q^{2n}P(q)\psi^2 (Rq),
\end{equation}
being $P(q)$ the power spectrum of the unfiltered field and $\psi (Rq)$ 
the Fourier transform of the circularly-symmetric linear filter.

\subsection{The presence of a local source}

Now, let us consider a Gaussian source (i. e. profile given by 
$\tau (x) = \exp ({- x^2/2R^2})$) embedded in the previous
background. Then,  
the expected number density of maxima per intervals $(x, x + dx)$,
$(\nu ,\nu + d\nu )$ and  
$(\kappa ,\kappa + d\kappa )$, given a source of amplitude $A$ in such
spatial interval, is given by (Barreiro et al.\cite{barreiro03})
\begin{equation}
n(\nu ,\kappa |\nu_s) =  \frac{n_b\,\kappa}{\sqrt{2\pi (1 - \rho^2)}}~
e^{- \frac{(\nu - \nu_s)^2 + (\kappa -
\kappa_s)^2 -  2\rho (\nu - \nu_s)(\kappa - \kappa_s)}{2(1 - \rho^2)}},
\label{nsource}
\end{equation}
where $\nu \in (-\infty ,\infty )$ and $\kappa \in (0,\infty )$, 
$\nu_s = A/\sigma_0$ is the normalized amplitude of the source and 
$\kappa_s = - A\tau_{\psi}^{\prime \prime}/\sigma_2$ is the
normalized curvature of the filtered source.
The last expression can be obtained as
\begin{equation}
\kappa_s = \nu_s y_s,~ 
y_s \equiv - \frac{\theta^2_m}{\rho} \tau_{\psi}^{\prime \prime},~
- \tau_{\psi}^{\prime \prime} = 2\int_0^{\infty}dq\,q^2\tau (q)\psi (Rq).
\end{equation}
We consider that the filter is normalized such that the amplitude of
the source is the same after linear filtering: $\int dx\,\tau (x)\Psi
(x; R, b) = 1$. 

\section{THE DETECTION PROBLEM}

We want to make a decision between filters based on \emph{detection}.
To make such a decision, we will focus on the following two 
fundamental quantities:
a) the number of spurious sources which emerge after the filtering 
and detection processes 
and b) the number of real sources detected. 
As seen in the previous section, these quantities are
properties of the Gaussian field and source that can be calculated
through equations (\ref{nbackground}) and (\ref{nsource}).
As we will see, the previous 
properties are not only related to the signal-to-noise-ratio gained in the filtering 
process but depend on the filtered momenta to 4th-order (in the 1D case), i. e. 
the amplification and the curvature of the source. 

\subsection{The region of acceptance}

Let us consider a local peak in the 1D data set characterized by the 
normalized amplitude and curvature $(\nu_s ,\kappa_s)$.
Let
$H_0: n.d.f. \ n_b(\nu ,\kappa ) \equiv n(\nu ,\kappa|0 )$ represents the \emph{null}
hypothesis, i. e. the local number density of background maxima, and 
$H_1: n.d.f. \ n(\nu ,\kappa |\nu_s )$ represents the \emph{alternative}
hypothesis, i. e. the local number density of maxima when there is a compact
source 
with normalized amplitude and curvature
$(\nu_s , \kappa_s = \nu_sy_s)$. 
Given the data $(\nu ,\kappa)$, we can associate to any region 
$\mathcal{R}_*(\nu,\kappa )$ two number densities $n_b^*$ and $n^*$ 
\begin{equation} \label{eq:nbstar}
n_b^* = \int_{\mathcal{R}_*} \,n(\nu ,\kappa |0)d\nu \,d\kappa, 
\end{equation}
\begin{equation} \label{eq:nstar}
n^* =  \int p(\nu_s ) d\nu_s \int_{\mathcal{R}_*} n(\nu ,\kappa |\nu_s )d\nu \,d\kappa.
\end{equation}
Then, $n_b^*$ is the number density of spurious sources, i. e. due to the
background,
expected inside the region $\mathcal{R}_*(\nu,\kappa )$, 
whereas $n^*$ is the number density of maxima 
expected in the same region of the $(\nu,\kappa)$ space
in the presence of a local source.
Hereinafter, we will call it the \emph{number density of detections}.
We will assume a Bayesian approach: at a concrete pixel we get the number of
source 
detections weighting with the \emph{a priori} probability
$p(\nu_s )$. $\mathcal{R}_*$ is called the \emph{acceptance} region.
 We remark that in
order to get the true number of real source detections such a number must be 
multiplied by the probability to have a source in a pixel in the original data
set.

We will assume a Bayesian Generalized Neyman-Pearson decision rule  
using number densities instead of probabilities: the
acceptance region $\mathcal{R}_*$ giving the highest number density of
detections $n^*$, 
for a given number density of spurious $n_b^*$, is the region (criterion for
detection)
\begin{equation}
\tilde{L}(\nu ,\kappa )\equiv \int_0^{\infty}d\nu_s p(\nu_s )L(\nu ,\kappa
|\nu_s)~
\equiv \frac{\int_0^{\infty} p(\nu_s )n(\nu ,\kappa |\nu_s )}{n(\nu ,\kappa
|0)}\geq L_*, 
\end{equation}
where $L_*$ is a constant. The proof follows the same approach as for the
standard Neyman-Pearson test.
Therefore, the decision rule is expressed by the likelihood 
ratio: if $\tilde{L}\geq L_*$ we decide that the signal is present, whereas 
if $\tilde{L}< L_*$ we decide that the signal is absent.

Once we have assumed the previous decision rule for detection, the region of
acceptance 
$\mathcal{R}_*$ is given by 
$\tilde{L}(\nu ,\kappa )\geq L_*$ or equivalently by the sufficient linear
detector 
\begin{equation}
\mathcal{R}_*:  \varphi (\nu ,\kappa )\geq \varphi_* ,
\end{equation}
where $\varphi_*$ is a constant and $\varphi$ is given by 
\begin{equation} 
\varphi (\nu ,\kappa )\equiv \frac{1 - \rho y_s}{1 - \rho^2}\nu +
\frac{y_s - \rho }{1 - \rho^2} 
\kappa,\ \ \ 
\mu \equiv \frac{{(1 - \rho y_s)}^2 }{1 - \rho^2}.
\label{phiymu}
\end{equation}
   We remark that
the assumed criterion for detection leads to a \emph{linear} detector
$\varphi$ (i. e. linear dependence on the threshold $\nu$ and curvature
$\kappa$).

\subsection{Spurious sources and real detections}

Once obtained the region of acceptance $\mathcal{R}_*$ in the previous
subsection, one 
can calculate the number density of spurious sources and the number density of 
detections as given by equations (\ref{eq:nbstar}) and (\ref{eq:nstar}).
\begin{eqnarray} 
n_b^* &=& \frac{n_b}{2}\left[ {\rm erfc}\left(
\frac{\varphi_*\sqrt{1-\rho^2}}
{\sqrt{2}(1-\rho y_s)}\right)~
+\sqrt{2}H y_s {\rm e}^{-H^2
\varphi_*^2} {\rm erfc}\left( -\frac{\sqrt{1-\rho^2}}{1-\rho y_s} y_s
H \varphi_* \right)\right], \nonumber \\
H&=&\sqrt{\frac{1-\rho^2}{2(1-2\rho y_s+y_s^2)}},
\label{naceptb}
\end{eqnarray}
\begin{eqnarray}  \label{eq:nstar2}
n^*  & = & \frac{n_b}{\sqrt{2\pi}}\frac{1 - \rho y_s}{(\mu +
y_s^2)\sqrt{1 - \rho^²}}~
\int_{\varphi_*}^\infty d\varphi I(\varphi )[1 + B(z)]e^{- \frac{(1
- \rho^2)\varphi^2}
{2{(1 - \rho y_s)}^2}},
\end{eqnarray}
\begin{equation}
z = \frac{y_s\varphi}{1 - \rho y_s}\sqrt{\frac{1 - \rho^2}{2(\mu + y^2_s)}}, 
\ \ 
I(\varphi ) = \int_0^{\infty}d\nu_s\,p(\nu_s)e^{\nu_s\varphi
-\frac{1}{2}\nu^2_s(\mu + y^2_s)},
\ \ 
B(x)\equiv \sqrt{\pi}xe^{x^2} \mathrm{erfc} (-\mathnormal{x}).
\end{equation}
Then, one can invert the equation for the number of spurious to get
$\varphi_* =\varphi_* (\frac{n^*_b}{n_b}; \rho , y_s) $ that allows to rewrite
the 
equation for the number of detections as $n^* = g(n_b^*; \theta_m, \rho ,y_s)$.

\section{ANALYTICAL AND NUMERICAL RESULTS}

\subsection{Point sources}

We will considerer as application the detection of compact sources
characterized by a  
Gaussian profile $\tau (x) = \exp (- x^2/2R^2)$, though the extension to other
profiles will be
considered in the future. Such a profile is physically and astronomically
interesting because
represents the convolution of a point source (Dirac $\delta$ distribution) with
a Gaussian beam.
 
\subsection{The matched filter (MF)}

By introducing a circularly-symmetric filter, $\Psi (x; R, b)$, 
we are going to express the conditions in order to obtain a
matched filter for the 
detection of the source $s(x)$ at the origin taking into account the
fact that the source is 
characterized by a single scale $R_o$. The following conditions are assumed: 
$(1) \  \langle w(R_o, 0)\rangle = s(0) \equiv A$, i. e. $w(R_o, 0)$ is an
\emph{unbiased} 
estimator of the amplitude of the source;
$(2)$ the variance of $w(R, b)$ has a minimum at the scale $R_o$,
i. e. it is an \emph{efficient} estimator
\begin{equation} \label{eq:mf}
\tilde{\psi}_{MF} = \frac{1}{2a}\frac{\tau (q)}{P(q)}.
\end{equation} 
\noindent This will be called \emph{matched} filter as is usual in
the literature.

For the case of a Gaussian profile for the source and a scale-free power 
spectrum given by $P(q)\propto q^{-\gamma}$, the previous formula leads to the 
following matched filter
\begin{equation} 
\tilde{\psi}_{MF} = \frac{1}{\Gamma (m)}x^{\gamma}e^{- \frac{1}{2}x^2},\ \ \
x\equiv qR,\ \ \  
m\equiv \frac{1 + \gamma}{2}.
\end{equation}

For the MF the parameters $\theta_c ,  \ \theta_m , \ \rho$ and the
curvature of the source $y_s$ are given by
\begin{equation}
\frac{\theta_m}{R} =  \frac{1}{\sqrt{1 + m}}  ,\ \ \ 
\rho = \sqrt{\frac{m}{1 + m}},\ \ \ 
y_s = \rho.
\end{equation}
We remark that the linear detector $\varphi (\nu ,\kappa )$ is reduced to
\begin{equation}
\varphi = \nu ,
\end{equation}
i.e. the curvature does not affect the region of acceptance for the MF and the sufficient 
detector $\varphi$ reduces to the plain thresholding detector. 

\subsection{The family of Matched-type filters (MTF)}

Let us modify the filtering scale as $\alpha R$ to introduce a family
of Matched-type filters. In particular, if we consider the white noise
case ($\gamma=0$), this family is given by
\begin{equation}
\label{eq:mtf}
\tilde{\psi}_{MTF} = \frac{1}{\sqrt{\pi}}\sqrt{\frac{1+\alpha^2}{2}}
e^{- \frac{1}{2}{(\alpha R q)}^2}.
\end{equation}
Therefore, this family allows one to filter at scales different from
the one of the source. Obviously, for $\alpha=1$ the usual MF is
recovered. 

The parameters characterizing the background and source are in this
case modified in the following way
\begin{equation}
\rho(\alpha)=\rho(\alpha=1),\ \ 
\theta_m(\alpha)=\alpha \theta_m(\alpha=1), \ \
y_s(\alpha)=\frac{2 \alpha^2}{1+\alpha^2} y_s(\alpha=1), \ \
\end{equation}
\begin{equation}
\label{eq:nualpha}
\nu(\alpha)= \sqrt{\frac{2\alpha}{1+\alpha^2}} \nu(\alpha=1), \ \
\kappa(\alpha)=\alpha^2 \sqrt{\frac{2\alpha}{1+\alpha^2}}
\kappa(\alpha=1).
\end{equation}

\subsection{Uniform distribution of point sources}

In this case, 
\begin{equation}
p(\nu_s ) = \frac{1}{\nu_c}, \ \ \ \nu\in [0, \nu_c]. 
\end{equation}
This allows one to obtain
\begin{equation}
I(\varphi ) = \sqrt{\frac{\pi}{2}}\frac{e^{u^2}}{\nu_c\sqrt{y^2_s + \mu }}
\left[ {\rm erf}(u) + {\rm erf} \left (\frac{\nu_c}{\sqrt{2}}\sqrt{y^2_s + \mu}
  -u \right) 
\right],\ \ u\equiv \frac{\varphi }{\sqrt{2(y^2_s + \mu )}}.
\end{equation}

\subsection{Theoretical results}

Hereinafter, we shall consider the case of white noise $(\gamma=0)$
for the background and a uniform distribution for the point sources
in an interval $\nu \in [0,\nu_c]$. $\nu_c$ has been chosen to produce
a threshold of 2 for the filtered field using the standard MF. The
corresponding thresholds for the MMF can be easily obtained using
equation (\ref{eq:nualpha}). We would like to point out that this
reflects a situation where we want to detect very weak sources, which
were below the $1\sigma$ level in the original unfiltered map.

\begin{figure}
   \begin{center}
   \begin{tabular}{c}
   \includegraphics[height=9cm]{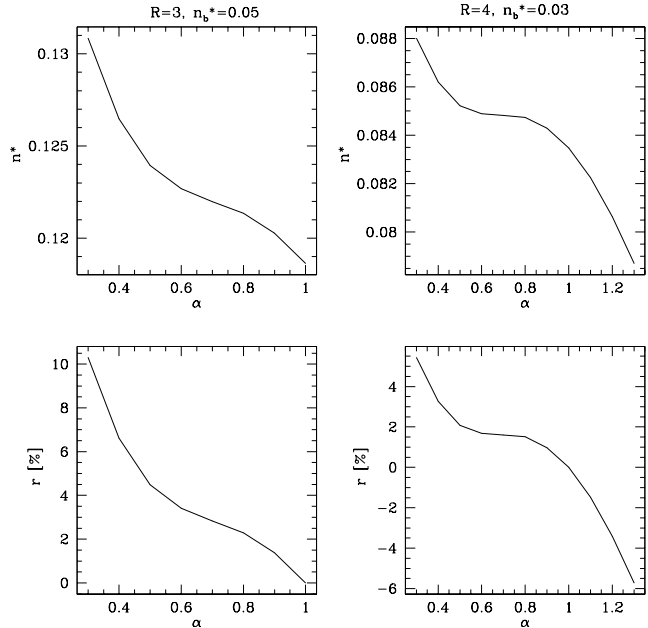}
   \end{tabular}
   \end{center}
   \caption[]{\label{fig:teoricas} The expected number density of
   detections (top panels) and the relative ratio to the standard MF
   (bottom panels) are given for
   the MTF for different values of the $\alpha$ parameter. Two cases
   are considered: $R=3$, $n_b^*=0.05$ (left figures) and $R=4$,
   $n_b^*=0.03$ (right figures).}
   \end{figure}

The region of acceptance $\mathcal{R}_*$ is defined by the sufficient linear
detector $\varphi$.
Let us consider equation (\ref{naceptb}), giving the number density of
spurious, then one can invert to get 
$\varphi_* (n^*_b)$. In Fig.~\ref{fig:teoricas} we show some results
for the cases $R$=3, $n_b^*=0.05$ and $R$=4, $n_b^*=0.03$. 
The two panels on the left correspond to the first case and the ones
on the right to the second one. The top figures show the expected
number density of detections versus the $\alpha$ parameter of the
filter. It is interesting to remark that such a number decreases with
$\alpha$. In fact, the largest number of detections are obtained for
$\alpha \simeq 0.3$ which corresponds approximately to filter at the
pixel scale. The relative ratio to the standard MF defined as
$r=(n^*(\alpha)/n^*(\alpha=1) - 1) \times 100$ is shown in the bottom
panels. For the first case, the ratio takes values up to $\simeq 10$ per cent
whereas for the second it goes up to $\simeq 5$ per cent. These
results clearly indicate that, under certain conditions, the standard
MF can be improved by simply modifying the scale of the filter. We
remark the importance of the curvature defining the acceptance
regions for the MTF.

\section{NUMERICAL SIMULATIONS: RESULTS}

\subsection{The simulations}

In order to test the previous ideas,
we have simulated a set of one-dimensional images containing a Gaussian background
characterized by a white noise power spectrum ($\gamma = 0$)
and point sources distributed in intensity following an uniform distribution
such as in eq. (24). Since we are interested in detecting very faint sources,
we set the upper-limit threshold cut $\nu_c = 2$
with respect to the MF.
Specifically, we simulated
a set of images with 4096 pixels each, with white noise dispersion unity (background image)
and then added a source with a Gaussian profile of FWHM=7 pixels, i.e. $R$ = 3, at the central pixel
of each image
(background+source image).
The size of the image is such that the addition of the source does not
modify the previous dispersion in a significant way. Then, each image
was filtered with the MTF given by equation (\ref{eq:mtf}).

For consistency, all the relevant quantities needed for the detection were estimated
directly from the images. The observables are the amplitude and curvature of
the maxima as well as the moments $\sigma_n^2$. From these observed quantities, it is
possible to calculate the parameters $\rho$ and $y_s$ of the filtered images and, therefore,
the value of the sufficient detector $\varphi$ (see equation \ref{phiymu}) for each  peak. 
The value $\varphi_*$ that defines the acceptance region
can be obtained from the images as well following a very simple procedure.
For any background maxima in the image, the detector $\varphi$  is calculated. Then all the
values of $\varphi$ are sorted. Once we fix the number of spurious sources $n^*_b$,
the value of $\varphi_*$ is given by the  $\varphi$ above which the number of
maxima found corresponds to $n^*_b$. Once defined the acceptance region, we can proceed to
apply the detector to any maxima where the existence of a source is suspected.
In our simulations, we looked at the central pixel of the images were the sources were
introduced and applied the detection criterion every time a maximum was found in that pixel.

\subsection{Results}

\begin{figure}
   \begin{center}
   \begin{tabular}{c}
   \includegraphics[height=9cm]{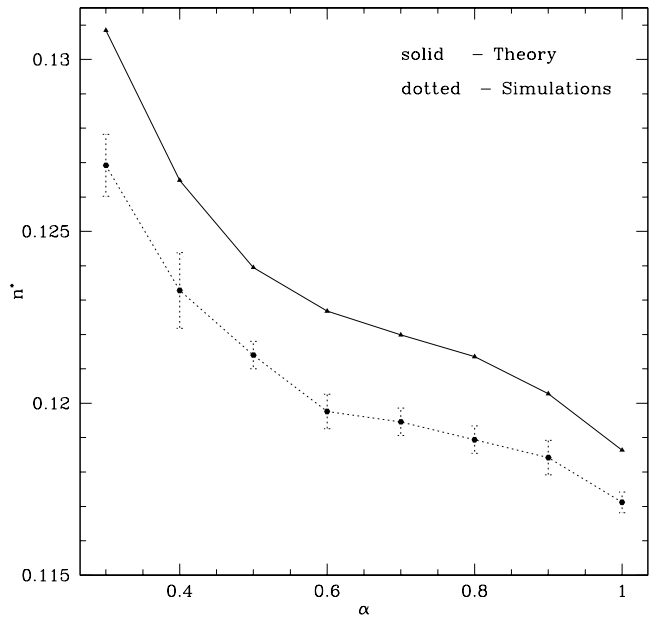}
   \end{tabular}
   \end{center}
   \caption[]{\label{fig:sims1} Number density of
   detections for the case  $R=3$, $n_b^*=0.05$ for different values of the 
   $\alpha$ parameter. The solid line represents the expected theoretical value
   whereas the dotted line is obtained through the numerical simulations.}
   \end{figure}

The results of the simulations are shown in figure~\ref{fig:sims1} for the case $R$ = 3,
$n^*_b$=0.05, where we represent the number of detections $n^*$ vs. the $\alpha$ parameter
of the filter. The dotted line represents the result obtained from the simulations whereas the solid
line is the theoretical result. We have done five simulations for each $\alpha$ value to estimate the 
error bars. In order to have 10000 images with a maximum of the background in the central pixel, for each of the
simulations we generate between $\simeq 50000$ and $\simeq 150000$ realizations (depending on $\alpha$). The two numbers reflect 
the fact that for lower $\alpha$ the same $n^*_b$ is achieved with a smaller number of realizations. We 
remark the agreement between the two results and also that
$\simeq 10$ per cent more sources are found filtering at the pixel scale ($\alpha \simeq 0.3$) than
with the standard MF ($\alpha = 1$).
Therefore, we have shown that in conditions of white noise and for weak sources
it is possible to outperform the number of detections given by the MF using
the detection criterion presented here and an appropriate MTF.
This clearly suggests the use of other optimal filters under certain
conditions, different from the MF, from the practical point of view, for the detection of weak sources.

This methodology is optimal for weak sources embedded in white noise. Regarding bright sources, the
amplification plays a  major role as compared to curvature entering the acceptance region. Therefore,
the use of the MF is well justified in this case.

Regarding the behavior of the detector with respect to the scale $R$, we found that
if $R$ is very low the sources become more and more point-like and they
are more easily mistaken with the fluctuations of the background. Therefore, in these cases
the curvature does not help to distinguish among them and the detector tends again to the thresholding
case which favors the use of the MF.

\section{CONCLUSIONS}

  The detection of compact sources on a background is a relevant
  problem for many fields, in
particular for Astronomy. Several detection techniques based on the use of
linear filters
thresholding-based detectors are standard. Here, we have considered an
  approach to the problem of detector design
based on a Bayesian generalization of the Neyman-Pearson rule that
  includes {\it a priori} information of the source distribution and the number densities of
  maxima (background and background plus source) to define the acceptance region.

  Our approach based on maxima includes both the amplitude and the
  curvature. Therefore, the chances of detection
do not depend only on the amplification of the sources
produced by the filtering but also on the filtered momenta up to the fourth order.
This determines in a strong way the designing of the linear filters
that are used to help the detection.
We have
  applied our technique to a
family of matched-type filters (MTF) by modifying the scale of the
  standard matched filter. We considered the very interesting case of
white noise to represent the background. As an example, we
  have considered a uniform
distribution of sources in the interval $0\leq \nu \leq 2 $ in the
  filtered field, i.e. weak sources. We have shown that the curvature
plays an important role defining the acceptance region and we have
  proven that the number 
of detections in the case of a filter with a scale similar to the pixel size
 beats the number of detections in the case of the standard MF.
This result has been tested with numerical
simulations for a uniform distribution and white noise.

  The ideas presented in this paper can be generalized: application to other
profiles (e. g. multiquadrics, exponential) and non-Gaussian backgrounds is
physically and astronomically interesting. The extension to include several images
(multi-frequency) is relevant. The generalization to two-dimensional data sets
(plane and spherical maps) and nD images is also very interesting. Finally the
application of our method to other fields is without any doubt. We are currently
doing research in some of these topics.

\pagebreak

\subsection{Acknowledgments} 
The authors thank Enrique Mart{\'\i}nez-Gonz\'alez and Patricio Vielva
for useful discussions. 
RBB thanks the Ministerio de Ciencia y Tecnolog\'\i a (MCYT) and the
Universidad de Cantabria for a Ram\'on y Cajal contract.
DH acknowledges support from the European Community's Human Potential Programme
under contract HPRN-CT-2000-00124, CMBNET. MLC thanks the MCYT for a
predoctoral FPI fellowship. We acknowledge partial support from 
the Spanish MCYT projects ESP2001-4542-PE and ESP2002-04141-C03-01
and from the EU Research Training Network `Cosmic Microwave Background
in Europe for Theory and Data Analysis'


\end{document}